\documentclass[preprint,showpacs,preprintnumbers,amsmath,amssymb]{revtex4}

\usepackage{epsfig}
\usepackage{graphicx}
\usepackage{dcolumn}
\usepackage{bm}

\let\jnfont=\rm
\def\NPB#1,{{\jnfont Nucl.\ Phys.\ B }{\bf #1},}
\def\PLB#1,{{\jnfont Phys.\ Lett.\ B }{\bf #1},}
\def\EPJC#1,{{\jnfont Eur.\ Phys.\ Jour.\ C }{\bf #1},}
\def\PRD#1,{{\jnfont Phys.\ Rev.\ D }{\bf #1},}
\def\PRL#1,{{\jnfont Phys.\ Rev.\ Lett.\ }{\bf #1},}
\def\MPLA#1,{{\jnfont Mod.\ Phys.\ Lett.\ A }{\bf #1},}
\def\JPG#1,{{\jnfont J.\ Phys.\ G}{\bf #1},}
\def\CTP#1,{{\jnfont Commun.\ Theor.\ Phys.\ }{\bf #1},}
\def\JHEP#1,{{\jnfont JHEP \ }{\bf #1},}
\def\NPPS#1,{{\jnfont Nucl.\ Phys.\ Proc.\ Suppl.\ }{\bf #1},}

\begin{document}

\preprint{\parbox{1.2in}{\noindent hep-ph/0609200}}

\title{\ \\[10mm]
Production of $h t \bar t$ and $h t \bar T$ in littlest Higgs model
with T-parity}

\author{\ \\[2mm] Lei Wang $^1$, Wenyu Wang $^1$, Jin Min Yang $^{2,1}$, Huanjun Zhang $^{3,1}$}

\affiliation{
$^1$ Institute of Theoretical Physics, Academia Sinica, Beijing 100080, China \\
$^2$ CCAST (World Laboratory), P.O.Box 8730, Beijing 100080, China \\
$^3$ Department of Physics, Henan Normal University, Xinxiang 453007,  China}

\begin{abstract}
In the littlest Higgs model with T-parity, which predicts a pair of
T-even and T-odd partners for the top quark, the top quark
interactions are altered with respect to the Standard Model
predictions and deviation will manifest in various top quark
processes. In this work we examine the effects in $ht\bar t$
productions at the ILC and LHC. We find that in the allowed
parameter space, the cross sections can be significantly deviated
from the Standard Model predictions and thus provide a good test for
the littlest Higgs model with T-parity. We also examine the new
production channel, the $ht \bar T$ or $hT \bar t$ production, at
the LHC, which give the same final states as $ht \bar t$ production
due to the dominant decay $T\to Wb$. We find that, compared with $ht
\bar t$ production, this new production channel can have a sizable
production rate for a $T$-quark below TeV scale. Such a production
will be counted into $ht\bar t$ events or possibly extracted from
$ht \bar t$ events, depending on if we can distinguish the $T$-quark
from the top quark from mass reconstructions.

\end{abstract}


\maketitle

\section{Introduction}

To solve the fine-tuning problem of the Standard Model (SM), the
little Higgs theory \cite{ref1} was proposed as a kind of
electroweak symmetry breaking  mechanism accomplished by a
naturally light Higgs sector. The Higgs boson remains light, being
protected by the approximate global symmetry and free from
one-loop quadratic sensitivity to the cutoff scale. The littlest
Higgs model \cite{ref2} provides an economical approach which
implements the idea of the little Higgs theory. Most of the
constraints from the electroweak precision tests  on little Higgs
models\cite{ref3} come from the tree-level mixing of heavy and
light mass eigenstates, which would require raising the mass of
the new particles to be much higher than TeV scale and thus
reintroduce the fine-tuning in the Higgs potential \cite{ref4}.
However, these tree-level contributions can be avoided by
introducing a discrete symmetry called T-parity \cite{ref5}. In
such a scenario, the top quark has a T-even partner  (denoted as
$T$) and a T-odd partner (denoted as $T_{-}$). As a result, the
top quark interactions are altered with respect to the SM
predictions, which will manifest in various top quark processes.
In this work, we will examine such effects in the associated $ht
\bar t$ productions at the LHC and ILC, and also study the $ht
\bar T$ and $hT \bar t$ productions at the LHC (due to the
heaviness of $T$-quark, $ht \bar T$ is beyond the threshold of the
ILC).

The reason for studying $ht \bar t$ production as a test of the
littlest Higgs model with T-parity is obvious.  Firstly, the large
top quark Yukawa coupling is speculated to be sensitive to new
physics and the $ht \bar t$ productions may be a sensitive probe
of the littlest Higgs model with T-parity. In this model the top
quark Yukawa coupling has a deviation from the SM prediction,
which will affect the $ht \bar t$ productions. Also the T-quark
can contribute to the $ht \bar t$ productions through its virtual
effects. Secondly, $ht \bar t$ production will be first searched 
at the LHC and can be precisely measured at the ILC
\cite{lhc-htt,ref8}. At the
ILC the top-quark Yukawa coupling can be measured with an accuracy
of about $5\%$ through the production of $ht\bar t$ \cite{ref9}
and the polarized beams can further improve the measurement
precision \cite{ref11}. The precision measurements of $ht \bar t$
production make it possible to unravel the new physics effects in
this process.

In addition, the new production channel at the LHC,  the $ht \bar T$
or  $hT \bar t$ productions, should also be considered since they give
the same final states as $ht \bar t$ production due to the dominant decay
$T\to Wb$.  As will be shown from our study,
compared with $ht\bar{t}$ production, this new production channel
can have a sizable production rate for a $T$-quark below TeV scale.
Such a production will be counted into $ht\bar t$ events or possibly
extracted from $ht \bar t$ events, depending on if we can distinguish
the $T$-quark from the top quark from mass reconstructions.

This work is organized as follows. In Sec. II we recapitulate the
littlest Higgs model with T-parity.
In Sec. III and Sec. IV
we study the $ht\bar{t}$ productions at the ILC and LHC, respectively.
In Sec. V we study the new $ht \bar T$ or $hT \bar t$  production channel
at the LHC. Finally, we give our conclusion at Sec. VI.

\section{About littlest Higgs model with T-parity}
Before our calculations we recapitulate the littlest Higgs model with
T-parity \cite{ref5,ref14}.
The gauge sector of this model can be simply obtained from the
usual littlest Higgs model  \cite{ref2}. T-parity acts as an automorphism which
exchanges the $[SU(2)\times U(1)]_{1}$ and $[SU(2)\times U(1)]_{2}$
gauge factors. Before electroweak symmetry breaking, the gauge boson mass
eigenstates have the simple form
\begin{equation}
W_{\pm}^{\alpha}=\frac{W_{1}^{\alpha}\pm W_{2}^{\alpha}}{\sqrt{2}},\
\ B_{\pm}=\frac{B_{1}\pm B_{2}}{\sqrt{2}},
\end{equation}
where $W_{j}^{\alpha}$ and  $B_{j}$ are $SU(2)_{j}$ and
$U(1)_{j}$(j=1,2) gauge fields. $W_{+}^{\alpha}$ and $B_{+}$ are the
$SM$ gauge bosons and have even T-parity, whereas $W_{-}^{\alpha}$ and
$B_{-}$ are additional heavy gauge bosons and have odd T-parity. After
electroweak symmetry breaking, the new mass eigenstates in the
neutral heavy sector will be a linear combination of
$W_{-}^{\alpha}$ and $B_{-}$ gauge bosons, producing $B_{H}$ and
$Z_{H}$. The $B_{H}$ is typically the lightest T-odd state and may
be a candidate of dark matter. Due to T-parity, the new gauge bosons
do not mix with the SM gauge bosons and thus generate no corrections
to precision electroweak observables at tree level. The top quark
sector contains a T-even and T-odd partner, with the
T-even one mixing with top quark and canceling the quadratic
divergence contribution of top quark to Higgs boson mass. The masses
of the T-even partner (denoted as $T$) and the T-odd partner
(denoted as $T_{-}$) are given by
\begin{eqnarray}
m_{T}\approx\frac{m_{t}f}{v}(r+\frac{1}{r}) , ~~~ m_{T_-}\approx
m_{T}s_{\lambda},
\end{eqnarray}
where $v$ is the electroweak breaking scale ($\approx 246$ GeV), $r=\lambda_1/\lambda_2$
with $\lambda_1$ and $\lambda_2$ are the coupling constants in the
Lagrangian of the top quark sector \cite{ref5,ref14,ref15}, and
$s_{\lambda}=1/\sqrt{1+r^{2}}$.

The mixing of $T$-quark with the top quark will alter the SM top
quark couplings and induce the couplings between $t$ and $T$
\cite{ref14,ref15}, which are given by
\begin{eqnarray}
V_{ht\bar{T}}&=&
         -m_{t}\left(\frac{s_{\lambda}^{2}}{f}P_{R}-\frac{c_{\lambda}}{s_{\lambda}v}P_{L}\right),\\
V_{Zt\bar{T}}^{\mu}&=&
     -\gamma^{\mu}\frac{e}{2S_{W}C_{W}}c_{\lambda}^{2}\frac{v}{f}P_{L}, \\
V_{Zt\bar{t}}^{\mu}&=&
    \gamma^{\mu}\frac{e}{S_{W}C_{W}}\left[\left(\frac{1}{2}-\frac{2}{3}S_{W}^{2}
    -\frac{c_{\lambda}^{4}}{2}\frac{v^{2}}{f^{2}}\right)P_{L}-\frac{2}{3}S_{W}^{2}P_{R}\right],\\
V_{ht\bar{t}}&=&
  -\frac{m_{t}}{v}\left(1-\frac{3+2r^{2}+3r^{4}}{4(1+r^{2})^{2}} \frac{v^{2}}{f^{2}}\right),
\label{htt}
\end{eqnarray}
where $P_{R,L}=(1\pm\gamma^{5})/2$ and $c_{\lambda}=r/\sqrt{1+r^{2}}$.
The $hZZ$ coupling involved in our calculations will also be different
from the SM coupling, which is given by
\begin{eqnarray}
V_{hZZ}^{\mu\nu}=\frac{2m_{Z}^{2}}{v}\left(1-\frac{1}{4}\frac{v^{2}}{f^{2}}\right)g^{\mu\nu} .
\end{eqnarray}
In the littlest Higgs model with T-parity, the T-quark can decay
into $Wb$, $ht$, $Zt$ and $B_{H}T_{-}$, among which the decay
$T\to Wb$ is the most important channel \cite{ref14,ref15,ref21}.
As shown in Fig. 12 of \cite{ref15}, $BR(T\to Wb)$ is over
46\% for $r=1.0$ and $500~{\rm GeV}\leq f\leq 2~{\rm TeV}$. When
$f$ is 500 GeV, $BR(T\to Wb)$ can be over 50\%. For comparison,
the subdominant decay $T\to Zt$ can have a branching ratio of
about $20\%$ at most in the parameter space for $r=1.0$ and
$500~{\rm GeV}\leq f\leq 2~{\rm TeV}$.

\section{Production of $h t \bar t$  at ILC}

Now we look at the process $e^{+}e^{-}\to t\bar th$ in the
littlest Higgs model with T-parity.
The Feynman diagrams are shown in Fig. 1.
In the SM it proceeds mainly through the $s$-channel $\gamma$ and $Z$
exchange diagrams with the Higgs boson radiated from the top quark,
as shown in Fig.1(a,b). Although a contribution can also come from
the diagram Fig.1(e) with the Higgs boson radiated from the gauge boson
$Z$, such a contribution is relatively small.
In the littlest Higgs model with T-parity we have additional diagrams Fig.1(c,d)
mediated by the $T$-quark. Due to the T-parity, other new particles, such as new heavy
gauge bosons $Z_{H}$ and $B_{H}$, do not participate in this process.

We calculate the cross section numerically by Monte Carlo
simulation. The cross section in the littlest Higgs model with T-parity
depends on two free parameters: the symmetry breaking scale $f$ and the ratio
$r=\lambda_1/\lambda_2$. Considering the electroweak precision
constraints \cite{ref13}, we vary them in the range $0.5\leq
r\leq5.0$ and $500$ GeV$\leq f\leq 2$ TeV. The SM parameters
involved are taken as $m_{t}=172.7$ GeV \cite{ref20}, $m_{h}=120$
GeV, $\alpha_{EW}=1/128.8$, $\sin^2\theta_{W}=0.2315$ and
$m_{Z}=91.187$ GeV \cite{ref16}.
\begin{figure}[tb]
\begin{center}\epsfig{file=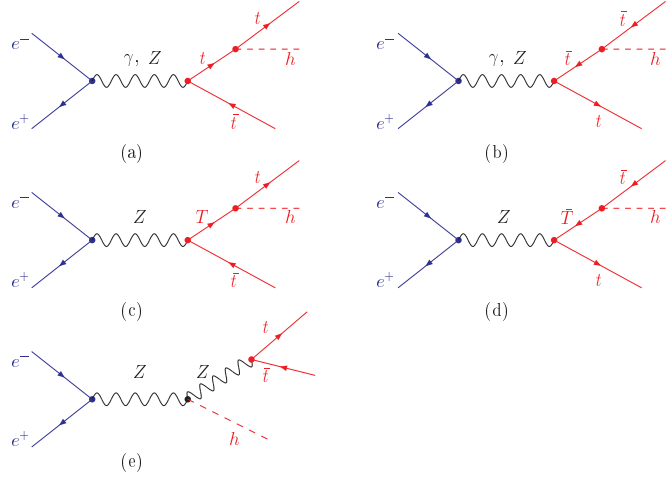,width=9cm} \end{center}
\vspace*{-1.0cm}
\caption{ Feynman diagrams for
          $e^{+}e^{-}\to t\bar t h$ in the littlest Higgs model with T-parity.}
\end{figure}
\begin{figure}[tb]
\begin{center}\epsfig{file=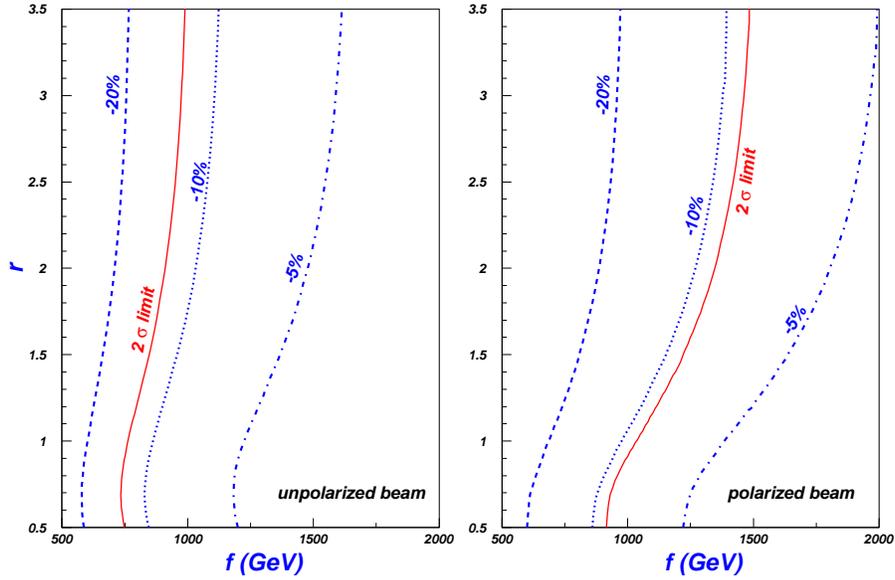,width=12cm}
\end{center}
\vspace*{-1.0cm}
\caption{The contours of the deviation from the SM
cross section $(\sigma-\sigma^{SM})/\sigma^{SM}$ for
$e^{+}e^{-}\rightarrow t\bar{t}h$ in the plane of
$r$ versus the symmetry breaking scale $f$. The solid curves are
         the $2\sigma$ statistical significance.}
\end{figure}

The c.m. energy is assumed to be $800$ GeV.
Considering the polarization of the initial electron and
positron beams, the cross section of $e^{+}e^{-}\rightarrow t\bar{t}h$
is given by \cite{ref17}
\begin{eqnarray}
\sigma&=&\frac{1}{4}\left[(1+p_{e})(1+p_{\bar{e}})\sigma_{RR}
 +(1-p_{e})(1-p_{\bar{e}})\sigma_{LL} \right. \nonumber \\
&& \left. +(1+p_{e})(1-p_{\bar{e}})\sigma_{RL}+(1-p_{e})(1+p_{\bar{e}})\sigma_{LR}\right],
\end{eqnarray}
where $\sigma_{RL}$ is the cross section for right-handed $e^-$ beam
($p_{e}=+1$) and left-handed $e^{+}$ beam  ($p_{\bar{e}}=-1$),
and other cross sections $\sigma_{RR}$, $\sigma_{LL}$ and $\sigma_{LR}$
are defined analogously.
As in \cite{ref11}, we assume $p_{e}=-0.8$ and $p_{\bar{e}}=0.6$ in our calculations.

In Fig. 2 we plot some contours for the deviation from the
SM cross section in the plane of $r$ versus the symmetry breaking scale $f$.
For comparison we also show the corresponding results for unpolarized
beams. We see that the polarized beams lead to more sizable deviation
and thus make the collider more powerful in probing such new physics
effects.
Fig. 2 shows that the contributions of this model decrease the SM
cross section in the allowed parameter space, and the magnitude of
such correction depends on the parameters $r$ and $f$. The
corrections are more sizable for lower values of the scale $f$, and
in a large part of the parameter space the contributions can alter
the SM cross section over $5\%$. When $f$ is lower than 1 TeV, the
corrections can be over $10\%$ in magnitude.

So far the electroweak precision data constrained the parameter
space of $r$ and $f$. But, as studied in \cite{ref13}, such
constraints depend on additional parameters, i.e., the masses of extra
T-odd fermions and the parameter $\delta_{c}$ whose value is dependent
on the details of the UV physics. Therefore, we did not show these
electroweak precision constraints in Fig. 2.

Another remarkable feature of our results is that the corrections are
very sensitive to the scale $f$, but not so sensitive to the parameter
$r$ when $r$ is larger than about 2, as shown in Fig. 2. This means that
we can use this process to determine or constrain the scale $f$ if
$r$ is large.

In Fig. 2  we also plotted the $2\sigma$ statistical significance,
obtained by assuming an luminosity of 1000 fb$^{-1}$ and an
efficiency of $10\%$ for events counting (due to kinematical cuts
and b-tagging, etc.). We see that a large part of parameter space is
within the $2\sigma$ statistical sensitivity. Of course, we should
note that some inevitable systematic error will worsen the probing
limits. Detector-dependent Monte Carlo simulations are necessary in
order to figure out the more practical probing limits.

Note that in the littlest Higgs model without T-parity, the new neutral gauge
bosons $Z_{H}$ and  $B_{H}$ can also contribute to the process
$e^{+}e^{-}\to t\bar th$ at tree-level via $s$-channel
resonances \cite{ref18}. In this case, the large values of $f$
required by the precision electroweak data suppress the
contributions of these new particles and, as a result, the T-quark
effects are very small. However, in the littlest Higgs model with
T-parity considered in this work, T-parity forbids the tree-level
contributions of the new gauge bosons $Z_{H}$ and $B_{H}$ to the
process since they are T-odd. Thus in this scenario only the T-quark
with even T-parity can contribute to the process at tree-level and
due to the relaxed constraint on $f$ (as low as 500 GeV is still
allowed), such T-quark effects may be sizable.
(However, we noticed that  there is an alternative
implementation of the T-parity \cite{ref6}, in which all new
particles which cancel the quadratic divergence of Higgs mass are
T-odd, including the top-quark sector. Thus, there is no
T-quark with even T-parity, and the T-quarks cannot contribute to
the process $e^{+}e^{-}\to h t\bar t$ at tree-level.)
\begin{figure}[tb]
\begin{center}\epsfig{file=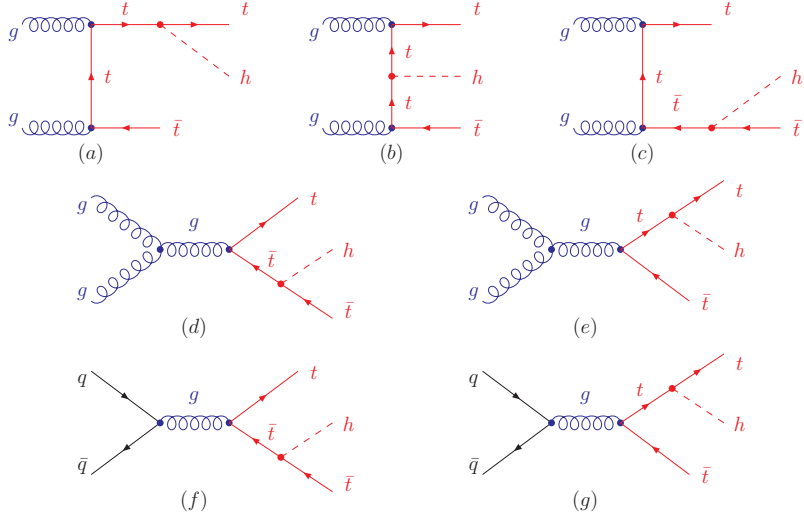,width=11cm}
\end{center}
\vspace*{-1.2cm}
\caption{The parton-level Feynman diagrams for $h t \bar t$ production
at LHC.  In the littlest Higgs model with T-parity, the  $h t \bar t$
vertex deviates from the SM value, as shwon in Eq.(\ref{htt}). The $u$-channel
diagrams by exchanging the two gluons in (a-c) are not shown here.}
\end{figure}
\begin{figure}[tb]
\begin{center}\epsfig{file=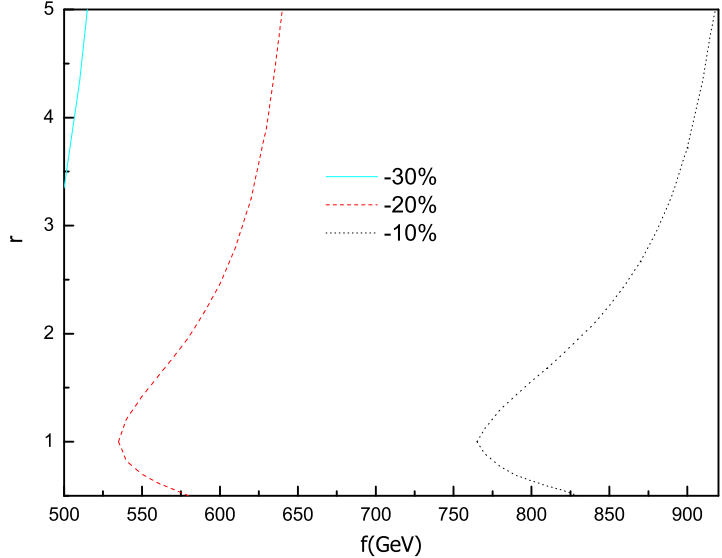,width=10cm}
\end{center}
\vspace*{-1.3cm}
\caption{The contours of deviation from the SM
cross section $(\sigma-\sigma^{SM})/\sigma^{SM}$ for the
process $pp \to h t\bar t +X$ at LHC.}
\end{figure}
\section{Production of $h t \bar t$  at  LHC}
The production of $h t\bar t$ at the LHC can proceed through $gg$
fusion or $q\bar{q}$ annihilation, as shown in Fig. 3.
In the littlest Higgs model with
T-parity the  $h t\bar t$ coupling is different from the SM prediction,
as shown in Eq.(\ref{htt}).
This will cause a correction to the production cross section
\begin{equation}
R=\frac{\sigma-\sigma^{SM}}{\sigma^{SM}}
 =\frac{V_{ht\bar{t}}^{2}-V_{ht\bar{t}}^{2}(SM)}{V_{ht\bar{t}}^{2}(SM)}.
\end{equation}
Here, $V_{ht\bar{t}} (SM)$ and $V_{ht \bar t}$ are the top-quark
Yukawa couplings in the SM and the littlest Higgs model with T-parity
\cite{ref14,ref15}, respectively.

Fig. 4 shows some contours for the deviation from the SM cross
section in the plane of $r$ versus the symmetry breaking scale $f$.
From this figure we see that the corrections decrease the SM cross
section in the allowed parameter space. The corrections are more
sizable for lower values of the scale $f$. In a large part of the
parameter space with $f<650$ GeV, the corrections can be over $20\%$
in magnitude.

\section{Productions of $h t \bar T$ and  $h T \bar t$ at  LHC}

Like $h t \bar t$ production, the production of $h t \bar T$ or $h T
\bar t$ can proceed through $gg$ fusion or $q\bar{q}$ annihilation
at the LHC, as shown in Fig. 5. In the littlest Higgs model with
T-parity, the T-quark can decay into $Wb$, $ht$, $Zt$ and
$B_{H}T_{-}$, among which the decay $T\to Wb$ is the most important
channel \cite{ref14,ref15,ref21}. Therefore, the final states of the
production $ht\bar{T}$ or $hT\bar{t}$ are same as $ht\bar{t}$
production. If we do not try to identify T-quark from top quark by
mass reconstruction, the productions $ht\bar{T}$ and $hT\bar{t}$
will be counted into $h t \bar t$ events. 

In Fig. 6 we plot the ratio
$\sigma(ht\bar{T}+hT\bar{t})/\sigma^{SM}(ht\bar{t})$ as a function
of T-quark mass. In our calculations we used the CTEQ5M patron
distribution functions \cite{ref23} with $Q=2m_{t}+m_{h}$ and
two-loop running coupling constant $\alpha_{s}(Q)$ with
$\alpha_{s}(m_{Z})=0.118$. From Fig. 6 we see that the ratio can
be over $10\%$ for $m_{T}$ below TeV scale. When $m_{T}$ is 700
GeV, the ratio can reach $40\%$. With the increase of $m_{T}$, the
production cross section becomes small because of the phase space
suppression.

Note that due to the large mass difference between $m_{T}$ and $m_t$,
we may try to extract the signal of $ht\bar{T}$ production from
$ht\bar t$ events by mass reconstructions. This is not easy since
it requires the mass reconstruction for both $t$ and $\bar t$.

\begin{figure}[tb]
\begin{center}\epsfig{file=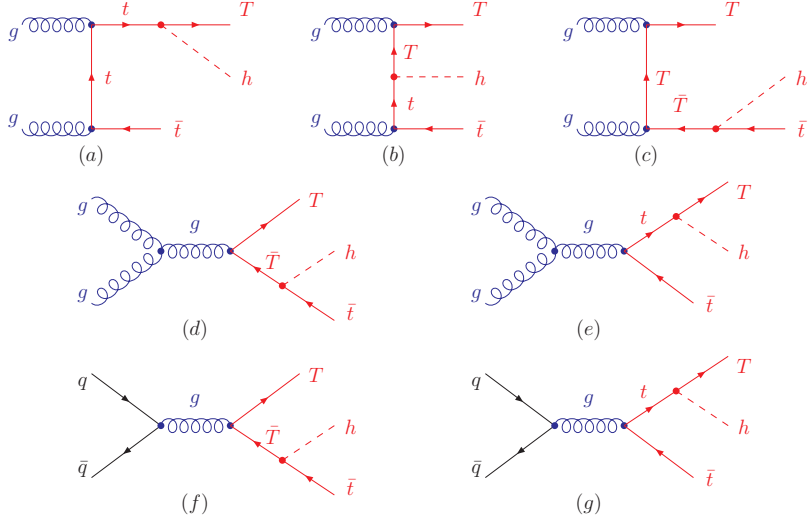,width=11cm}
\end{center}
\vspace*{-1.0cm}
\caption{The parton-level Feynman diagrams for $h T \bar t$ production
at LHC in the littlest Higgs model with T-parity.  The $u$-channel
diagrams by exchanging the two gluons in (a-c) are not shown here.}
\end{figure}
\begin{figure}[tb]
\begin{center}\epsfig{file=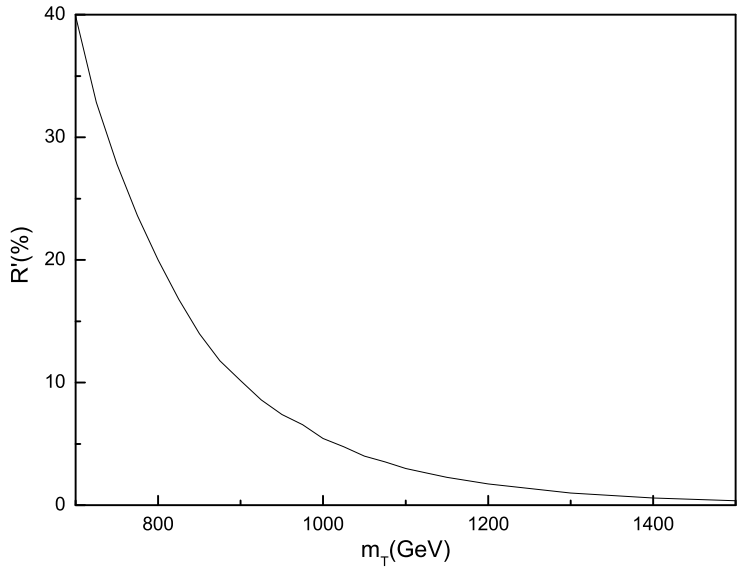,width=11cm}
\end{center}
\vspace*{-1.2cm} \caption{The ratio
$R'=\left[\sigma(ht\bar{T}+hT\bar{t})\right]/\sigma^{SM}(ht\bar{t})$
at LHC as a function of $m_{T}$ for $r=1.0$.}
\end{figure}

Given the analyses in both this section and the preceding section,
we would like to remark on the overall impact of the modified
cross sections for the Higgs discovery at the LHC. As shown in
\cite{Drollinger}, the $ht\bar{t}$ production channel will be hard 
to be observed at the LHC. As shown in Sec. IV, the contribution of 
the littlest Higgs model with T-parity can decrease the SM $ht\bar{t}$ 
cross section by $20\%$, which thus makes the observation of this
production channel even harder.  
But, at the same time, the new channels of
$hT\bar t$ and $ht\bar T$ production may open up.
As shown in Fig. 6, for $700~{\rm GeV}<m_{T}<800~{\rm
GeV}$ the production of $hT\bar t$ and $ht\bar T$ can have a cross
section of $20\%\sim 40\%$ with respect to the SM $ht\bar t$ cross
section. Considering the heaviness of the $T$-quark, the
production of $ht\bar{T}+hT\bar{t}$ may have less background than
$ht\bar{t}$ production, and thus this new channel
may likely be observable at the LHC.

\section{Conclusion}
We studied top quark pair production associated with a light Higgs
boson as a test of the littlest Higgs model with T-parity at the
ILC and LHC. For the production of $h t\bar{t}$ at the ILC, we
found that in a large part of the allowed parameter space the
cross section can deviate from the SM prediction by over $10\%$
and thus may be observable. Also, we found that the polarized
beams lead to more sizable deviation and thus make the ILC more
powerful in probing such effects. For the production of $h
t\bar{t}$ at the LHC, we found that in a large part of the
parameter space the deviation from the SM cross section can be
over $20\%$. For the new
production channel of $ht\bar{T}$ or $hT\bar{t}$, 
we found that their cross section can be over 10\% of the SM
$ht\bar{t}$ production for $m_{T}$ below TeV scale. 

\vspace*{0.6cm} \vspace{0.5cm} \noindent{\bf Acknowledgments:} L.
W. thanks C.-X. Yue and J. J. Cao for discussions. This work was
supported by National Natural Science Foundation of China (NNSFC)
under No. 10475107.

\end{document}